\documentclass[pra,twocolumn,amsmath]{revtex4}

\usepackage{graphicx}

\begin{document}

\title{Coherent Control and Entanglement in the Attosecond
Electron Recollision Dissociation of D$_2^+$}

\author{Michael Spanner}
\author{Paul Brumer}
\affiliation{Chemical Physics Theory Group, Department of Chemistry, and Center for Quantum Information and Quantum Control, University of Toronto, Toronto, M5S 3H6 Canada}

\date{\today}

\begin{abstract}
We examine the attosecond electron recollision dissociation 
of D$_2^+$ recently demonstrated experimentally
[H. Niikura {\it et al.}, Nature (London) {\bf 421}, 826 (2003)]
from a coherent control perspective.
In this process, a strong laser field incident on D$_2$ 
ionizes an electron, accelerates the electron in the laser field 
to eV energies, and then drives the
electron to recollide with the parent ion, causing
D$_2^+$ dissociation.
A number of results are demonstrated.
First, a full dimensional Strong Field Approximation (SFA) model
is constructed and shown to be in agreement with the original 
experiment.  This is then used to rigorously demonstrate that
the experiment is an example of coherent pump-dump control.
Second, extensions to bichromatic coherent control are proposed
by considering dissociative recollision of molecules prepared in a
coherent superposition of vibrational states.
Third, by comparing the results to 
similar scenarios involving field-free attosecond 
scattering of independently prepared D$_2^+$ and electron 
wave packets, recollision dissociation is shown to provide 
an example of wave-packet coherent control of 
reactive scattering.  Fourth, this analysis 
makes clear that it is the temporal correlations between the 
continuum electron and D$_2^+$ wave packet, and not 
entanglement, that are crucial 
for the sub-femtosecond probing resolution demonstrated in the
experiment.  This result clarifies some misconceptions regarding 
the importance of entanglement in the recollision probing of D$_2^+$.
Finally, signatures of entanglement 
between the recollision electron and the atomic fragments,
detectable via coincidence measurements, are identified.
\end{abstract}

\maketitle

\section{Introduction}

Laser-based quantum control techniques can be loosely characterized
by two general control ideologies: scenarios based
in a time-domain perspective, such as pump-dump \cite{PumpDump} control, 
and scenarios based in the energy domain such as 
multi-path interference \cite{BrumerShapiro} and 
STIRAP \cite{STIRAP} control.  
The quantum control of reactive scattering also fits
these paradigms, where examples of multi-path interference
(energy-domain) \cite{BrumerScattering}
and wave packet (time-domain) \cite{Me}
schemes for coherently controlled chemical reactivity can be found.
In this article, the coherent control of attosecond electron
recollision dissociation is studied using both pump-dump 
and multi-slit interference schemes, and we find that this
process offers the first experimental demonstration of 
wave-packet coherent control 
of reactive scattering \cite{Me}.

The recollision experiment that is the focus of this 
paper proceeds as follows.
Using femtosecond lasers of intensities around $10^{14}-10^{15}$ W/cm$^2$,
one ionizes an atom or molecule near a peak of the
instantaneous electric field, accelerating the liberated electron 
in the laser field, and causing the electron to  
recollide with the parent ion \cite{Plasma}.
The ionization/recollision process occurs in less than one
cycle of the laser field, and repeats every cycle when the
peak electric field is large enough to cause significant ionization.  
For Ti:sapphire laser systems (800 nm) typical of recollision experiments, 
one optical period is 2.6 fs and the recollision of the 
continuum electron wave packet with the core lasts about 500 asec.
A number of processes can occur upon recollision.  The continuum
electron can, for example, recombine with the ion while emitting its 
excess energy as a burst of XUV radiation \cite{HHG}, or scatter
elastically thereby taking a sub-femtosecond electron diffraction 
image of the ion \cite{Diffraction}.  Alternatively, as studied here,
the molecule can undergo recollision-induced dissociation 
\[ {\rm D}_2  \longrightarrow
 {\rm D}_2^+ + {\rm e}^- \longrightarrow
{\rm D} + {\rm D}^+ + {\rm e}^-, \] 
where the strong field ionizes D$_2$ in the first step.
In the second step, the continuum electron 
excites the bound electron from the $\Sigma_g$ bonding
state to the $\Sigma_u$ antibonding state thereby dissociating
the molecule.  This process
has been demonstrated experimentally, and used to probe the
vibrational motion of the D$_2^+$ nuclear wave packet
following the initial ionization of D$_2$ on sub-femtosecond 
time scales \cite{CorkumRecollision}.

Here, we consider the dissociative recollision
process from a coherent control perspective.  First,
a full dimensional quantum model, based on the Strong Field 
Approximation (SFA), is constructed.  The model is then validated by 
successfully simulating the experiment, which is 
shown to be an example of pump-dump control, consistent with
a perspective long-held by the experimental NRC group \cite{foot1}.
Second, the scenario is extended to {\it bichromatic} 
coherent control using an initial superposition of 
D$_2$ vibrational states.  Considerable control is demonstrated,
motivating future experimental studies.  Third, 
field-free scattering of D$_2^+$ and e$^-$ wave packets is
studied in order to connect the observed control of 
dissociative recollision with the recently constructed 
theory of wave-packet coherent control of reactive 
scattering \cite{Me}.  Fourth,
the role of entanglement, previously suggested to be connected to 
the vibrational probing 
scenario \cite{CorkumRecollision,Entangle1,EntangleMarkus}, is 
investigated.  Specifically, by comparing the vibrational probing 
scenario to similar scenarios using field-free non-entangled 
scattering states we demonstrate that it is the temporal
correlations between the scattering wave packets, and not 
entanglement, that allow for probing and control in dissociative 
recollision.   Nevertheless, entanglement is still present, and
we conclude the paper by identifying its signatures, detectable
in coincidence measurements.

For coherent control, these 
results are of interest because they identify new experimentally 
accessible examples of both bichromatic control and 
wave-packet control of reactive scattering.
For strong field recollision, these
results are important  because
they clarify the role of entanglement 
in recollision-based probing techniques \cite{CorkumRecollision},
and motivate new strong field recollision control experiments.


\section{Methodology}

\subsection{Strong Field Recollision in SFA}

The SFA, a well-known method in strong field 
physics \cite{Reiss}, is here developed for the dissociative 
recollision scenario.  We utilize the SFA in the length gauge,
and work in the single active electron approximation in that
only the action of the laser on the ionized electron is considered,
while the bound electron and nuclei do not interact directly
with the strong field.  In general, there could also be strong 
laser-induced coupling between the bound-electronic states
leading to strong field molecular effects (e.g. bond-softening, 
enhanced ionization) that would affect the nuclear states.
However, the dissociative recollision experiments being
modeled used an angle-limited detection scheme to 
selectively measure D$^{+}$ fragments coming 
predominantly from molecules aligned perpendicular to the
laser field \cite{CorkumRecollision}.  Strong field 
molecular effects are minimized for this geometry, hence 
justifying their exclusion from the analysis.
Note that the language below is specific to D$_2$, but the
formalism is completely general.

The exact solution for the wave function 
propagation can be written in the form
\begin{eqnarray}\label{EqPartExact}
	|\Psi(t)\rangle &=& 
	\widehat U(t,t_0)|\Psi_i\rangle  
	\\ \nonumber
	&=& -i\int^t_{t_0} dt'
	\widehat U(t,t')
	\widehat V_L(t') 
	e^{-i (t'-t_0) \widehat H_0}|\Psi_i\rangle 
	\\ \nonumber
	&+& e^{-i(t-t_0) \widehat H_0}|\Psi_i\rangle 
\end{eqnarray}
where $\widehat H_0$ is the field free Hamiltonian,  
$\widehat V_L(t) = {\bf E}(t) \cdot \widehat {\bf r}$
is the laser-matter interaction for the electric field
${\bf E}(t)$ and the electron position $\widehat {\bf r}$.  
The operator $\widehat U(t,t_0)$ 
is the full propagator, defined by
\begin{equation}
	i \frac{\partial} {\partial t} \widehat U(t,t_0)
	= \widehat H(t) \widehat U(t,t_0), 
	\:\:\:\: \widehat U(t_0,t_0) = \widehat I,
\end{equation}
where $\widehat H(t) = \widehat H_0 + \widehat V_L(t)$ is the
system+laser Hamiltonian, and $\widehat I$ is the identity
operator.  Note that all equations
are written in atomic units, $\hbar = m_e = e = 1$.
The vector potential ${\bf A}(t)$ is chosen to be
\begin{equation}\label{EqA}
	{\bf A}(t) = -\frac{{\cal E}_0}{\omega}\sin(\omega t) \hat x,
\end{equation}
thus giving the electric field
\begin{equation}\label{EqE}
	{\bf E}(t) = -\frac{\partial {\bf A}(t)}{\partial t} =
	{\cal E}_0\cos(\omega t) \hat x.
\end{equation}
The initial state $|\Psi_i\rangle$ is assumed to be in 
an eigenstate of the field-free system.
The last term on the right hand side of Eq. (\ref{EqPartExact}) 
describes the evolution of
the unperturbed component of the initial wave function.  
It does not contribute to the dissociative recollision channel 
and is therefore dropped from subsequent expressions. 
Expanding now the propagator $\widehat U(t,t')$ 
appearing in Eq. (\ref{EqPartExact}) in a manner 
similar to Eq. (\ref{EqPartExact}), but
this time expanding in the electron-electron interaction 
$\widehat V_{ee}$, yields
\begin{eqnarray}\label{EqRecollisionWave}
	|\Psi(t)\rangle &=& -\int^t_{t_0} dt' \int^t_{t'}dt'' \:
	\widehat U(t,t'')
	\widehat V_{ee}(t'') \widehat U(t'',t')\times \nonumber \\ 
	&&
	\widehat V_L(t') e^{-i(t'-t_0)\widehat H_0}
	|\Psi_i\rangle,
\end{eqnarray}
where another benign term, this time related to direct ionization 
{\it without} recollision, was again dropped since it also does not 
contribute to the process under consideration.

Before applying the SFA to this exact expression for the dissociative
recollision, Eq. (\ref{EqRecollisionWave}) is first connected to 
the physical picture of ionization/propagation/recollision \cite{Plasma}
introduced above.  In particular, consider the integrand of Eq. 
(\ref{EqRecollisionWave}) from right to left.  First, the 
initial state $|\Psi_i\rangle$ propagates field-free until
the time $t'$, when it gets a 'kick' from the
laser field $\widehat V_L(t')$.  This is followed by propagation 
using the full Hamiltonian for a time $(t''-t')$ after which
the wave function is affected by the electron-electron
interaction $\widehat V_{ee}$.  The wave function then propagates
to the observation time $t$.  Thus, the partitioning used in
expanding the full propagators has yielded
a propagation sequence that closely resembles the physics:
The active electron first sits in the ground state until 
being ionized by the laser.  This qualitatively matches the
operators in Eq. (\ref{EqRecollisionWave})  to the right of,
and including, $\widehat V_L(t')$.
Following ionization, 
the electron then oscillates in the continuum driven by the 
laser field until it recollides with the core.  These steps 
qualitatively match the effect of the operators 
$\widehat V_{ee}(t'') \widehat U(t'',t')$.

With the formalism firmly connected to the physical picture, 
the SFA can now be applied.
The central approximation introduced by the SFA is to neglect the
interaction between the ion and the continuum electron
in the intermediate propagator 
$\widehat U(t'',t')$ where the electron
propagates far from the core before returning to the ion,
and also in the final propagator 
$\widehat U(t,t'')$.  This means that
once the active electron has been ionized, it 
acts as a completely free electron oscillating in the laser field.
The only deviation from this idealization arises due to
the electron-electron interaction $\widehat V_{ee}(t'')$.
From the point of view of scattering theory, this means
that after ionization the continuum electron is effectively being
treated by a Born-type approximation in the resulting
recollision event, albeit a Born approximation dressed by 
the strong laser field that drives the electron recollision. 

In order to carry out the SFA, complete basis sets 
consisting of the field-free molecular states for the nuclear
component and plane waves for the active electron
are inserted between the operators appearing in 
Eq. (\ref{EqRecollisionWave}) to give
\begin{widetext}
\begin{eqnarray}\label{EqExpandedBasis}
	|\Psi(t)\rangle &=& - \sum_{{\bf n},{\bf n}',J} \int dE
	\int^t_{t_0} dt' \int^t_{t'}dt''
	\int d {\bf k} \int d{\bf k}' \int d{\bf k}'' \\ \nonumber
	&\times& 
	\widehat U(t,t'')
	|\phi^{(u)}_{E,J};{\bf k}'',t''\rangle \:
	\langle \phi^{(u)}_{E,J};{\bf k}'',t''|
	\widehat V_{ee}(t'')
	|\phi^{(g)}_{{\bf n}'};{\bf k}',t''\rangle \nonumber \\ &\times& 
	\langle\phi^{(g)}_{{\bf n}'};{\bf k}',t''|
	\widehat U(t'',t')
	|\phi^{(g)}_{\bf n};{\bf k},t'\rangle 
	 \langle\phi^{(g)}_{\bf n};{\bf k},t'|
	\widehat V_L(t') e^{-i E_i (t'-t_0)}|\Psi_i\rangle \nonumber,
\end{eqnarray}
where $E_i$ is the energy of the initial state presently
assumed to be an eigenstate of $\widehat H_0$ (superposition
states will be considered below),
 $|\phi^{(g)}_{\bf n}\rangle$ are the bound nuclear states on
the $\Sigma_g$ surface with quantum numbers ${\bf n} = (\nu,J,m)$, 
$|\phi^{(u)}_{E,J}\rangle$ are the continuum states with energy $E$ 
and and angular momentum $J$ on the $\Sigma_u$ surface, the 
electronic continuum (plane wave) states are
\begin{equation}\label{EqCanonicalBasis}
	\langle{\bf r}|{\bf k},t\rangle = (2\pi)^{-(3/2)} 
	e^{i[{\bf k}+{\bf A}(t)] \cdot {\bf r}},
\end{equation}
and ${\bf k}$ is the canonical momentum of the free electron
oscillating in the laser field.  Within the 
above outlined assumptions, and after projecting
onto a particular final state $|\phi^{(u)}_{E,J};{\mathbf k}_f,t\rangle$, 
Eq. (\ref{EqExpandedBasis}) reduces to
\begin{eqnarray}
	\langle\phi^{(u)}_{E,J};{\mathbf k}_f,t|\Psi(t)\rangle &=& - \sum_{\bf n}
	\int^t_{t_0} dt' \int^t_{t'}dt''
	\int d{\mathbf k} \\ &&
	e^{-(i/2)\int^t_{t''} d\tau [{\mathbf k}_f+{\mathbf A}(\tau)]^2}
	e^{-iE(t-t'')} 
	\langle \phi^{(u)}_{E,J};{\mathbf k}_f,t''|
	\widehat V_{ee}(t'')
	|\phi^{(g)}_{\bf n};{\mathbf k},t''\rangle  \nonumber
	\\ && 
	e^{-(i/2)\int^{t''}_{t'} d\tau [{\mathbf k}+{\mathbf A}(\tau)]^2}
	e^{-iE_{\bf n}(t''-t')}e^{-i E_g (t'-t_0)} 
	 \langle\phi^{(g)}_{\bf n};{\mathbf k},t'|
	\widehat V_L(t') |\Psi_i\rangle ,\nonumber
\end{eqnarray}
where $E_{\bf n}$ is the energy of the nuclear state ${\bf n}$,
and the propagation of the continuum electron 
was handled using the Gordon-Volkov solutions \cite{volkov}
\begin{eqnarray}\label{EqPreStation}
	\langle {\mathbf k',t''}| e^{-i\int^{t''}_{t'} \widehat H_V(\tau)d\tau }
	|{\mathbf k,t'}\rangle &=& 
	\langle {\mathbf k',t''}| 
	e^{-(i/2)\int^{t''}_{t'}d\tau [{\mathbf k}+{\mathbf A}(\tau)]^2} 
	|{\mathbf k,t''}\rangle  \\ \nonumber
	& =&
	e^{-(i/2)\int^{t''}_{t'} d\tau [{\mathbf k}+{\mathbf A}(\tau)]^2}
	\delta({\mathbf k} - {\mathbf k'}),
\end{eqnarray}
which are exact solutions for a free electron oscillating in a laser field.
The remaining integrals in Eq. (\ref{EqPreStation}) are 
evaluated using the stationary phase method,
keeping the stationary points on the real time axis, yielding
\begin{eqnarray} \label{EqAfterStation}
	\langle\phi^{(u)}_{E,J};{\bf k}_f,t|\Psi(t)\rangle &=& - 
	\sum_{\bf n} 
	\sum_{j=S,L} 
	e^{i\varphi({\bf k}_0,t^{(j)}_b,t^{(j)}_c) } 
	a_c(t_c,t_b) a_p(I_{p,{\bf n}},t^{(j)}_c-t^{(j)}_b)
	a_i(I_{p,{\bf n}},t^{(j)}_b)
	\nonumber \\ & \times &
	\langle \phi^{(u)}_{E,J};{\bf k}_f,t^{(j)}_c|
	\widehat V_{ee}(t^{(j)}_c)
	|\phi^{(g)}_{\bf n};{\mathbf k}_0,t^{(j)}_c\rangle  
	 \langle\phi^{(g)}_{\bf n};{\mathbf k}_0,t^{(j)}_b|
	\widehat V_L(t^{(j)}_b) |\Psi_i\rangle, 
\end{eqnarray}
where 
\begin{equation}
	\varphi({\bf k}_0,t_b,t_c) = 
	-\frac{1}{2}\int^t_{t_c} d\tau [{\bf k}_f+{\bf A}(\tau)]^2
	-E(t-t_c)
	-\frac{1}{2}\int^{t_c}_{t_b} d\tau [{\bf k}_0+{\bf A}(\tau)]^2
	-E_{\bf n}(t_c-t_b)
	-E_i (t_b-t_0),
\end{equation}
\begin{equation}
	a_c(t_c,t_b) =
	e^{i\pi/4}(2\pi)^{1/2}\left|
		-{\bf E}(t_c)[{\bf k}_f+{\bf A}(t_b)]
	+{\bf E}(t_b)[{\bf A}(t_b)-{\bf A}(t_c)]
	\frac{\partial t_b}{\partial t_c}
	+ I_{p,{\bf n}}
	\frac{\partial^2 t_b}{\partial t^2_c}
	\right|^{-\frac{1}{2}}
\end{equation}
\begin{equation} \label{EqProp}
	a_p(I_{p,{\bf n}},t_c-t_b) =
	e^{-i3\pi/4}\left[ 
	\frac{2\pi}{|t_c-t_b|}\right]^{3/2}
\end{equation}
\begin{eqnarray}\label{EqAiry}
	a_i(I_{p,{\bf n}},t_b) =
	\sqrt{\pi}\left[\frac{2}
	        {I_{p,{\bf n}}\left|{\mathbf E}(t_b) \right|^2}\right]^{1/4}
	\exp\left[-\frac{1}{3}
	          \frac{(2I_{p,{\bf n}})^{3/2}}
	{\left|{\mathbf E}(t_b) \right|}\right],
\end{eqnarray}
the state-to-state ionization potentials are
$I_{p,{\bf n}} = E_{\bf n} - E_i$,
the stationary phase points for the
time of birth ($t_b$), time of collision ($t_c$), and
momentum at birth (${\bf k}_0$)  are given by
\begin{subequations} \label{EqFinalStation}
\begin{eqnarray}
	{\bf k}_0 = -{\bf A}(t_b) &=&  \frac{{\cal E}_0}{\omega}
	\sin(\omega t_b) \hat x \\
	\omega*(t_c-t_b)*\sin(\omega t_b) &=& 
	\cos(\omega t_b) - \cos(\omega t_c) \\
	\frac{1}{2} \left|{\bf k}_f -
	\frac{{\cal E}_0}{\omega}\sin(\omega t_c)\hat{\bf z}\right|^2 
	&=&
	 \frac{1}{2} \frac{{\cal E}^2_0}{\omega^2}\left[\sin(\omega t_b)  
	-\sin(\omega t_c)\right]^2  -D_{E,{\bf n}}
	- I_{p,{\bf n}} \frac{\partial t_b}{\partial t_c}
\end{eqnarray}
\end{subequations}
\end{widetext}
and $D_{E,{\bf n}} = E -  E_{\bf n}$ is the state-dependent 
dissociation energy.  The derivatives of $t_b$ with respect to $t_c$
are defined through Eq. (\ref{EqFinalStation}b).
For each ${\bf k}_f$, 
Eqs. (\ref{EqFinalStation}) admit two solutions, the long (L) and
short (S) trajectories, where 
$t^{(S)}_c-t^{(S)}_b < t^{(L)}_c-t^{(L)}_b$.  
The sum over $j$ in Eq. (\ref{EqAfterStation}) accounts for 
these two solutions. 

In principle, Eq. (\ref{EqAfterStation}) is the final 
stationary phase SFA amplitude for the dissociative 
recollision process.  However, some additional 
simplifications relevant to the present work are in order.
First, note that while the amplitude Eq. (\ref{EqAfterStation})
correctly captures much of the essential physics of 
recollision problems, such as the qualitative dependence of 
the final yields on laser frequency and intensity, it 
typically underestimates absolute yields by 1 to 2
orders of magnitude.  This is due to the neglect the Coulomb
potential during the ionization step.  However, methods
are available to correct the ionization rates, on sub-cycle
time scales, by adding the Coulomb potential in 
some approximate way \cite{Coulomb}.  Second, $a_c(t_c,t_b)$
in Eq. (\ref{EqAfterStation}) has divergences.
These divergences are artificial in the sense that the 
stationary phase method is simply not valid at these 
points. Rather, one should use the uniform approximation 
to treat these regions.  When treated correctly,
$a_c(t_c,t_b)$ is a
slowly-varying prefactor (as a function of the
final state parameters) to the amplitude Eq. (\ref{EqAfterStation})
and affects little but the absolute yield.
Since the present study is focused on 
building a novel perspective of this process based upon  
ideas in coherent control, we choose to avoid additional 
computational complications, and simply drop the divergent 
term, recognize that the absolute value is
nonquantitative and accept that this model correctly captures
the dependence of the yields on the laser and initial state
parameters, but gives incorrect absolute yields. Similarly, 
sub-cycle corrections to the ionization rate are not included.
Finally, dividing 
Eq. (\ref{EqFinalStation}c) by the maximum value of the 
vector potential ${\cal E}_0/\omega$ shows that
the term containing $\partial t_b/\partial t_c$ is proportional to
$I_p/(2U_p) = \gamma^2$, where $\gamma$ is the Keldysh parameter
\cite{Keldysh}. 
In the tunneling ionization regime, $\gamma$ is assumed to be a 
small parameter, and thus the $\partial t_b/\partial t_c$ 
term in Eq. (\ref{EqFinalStation}c) can be neglected.  
In the calculations present below, this term was neglected.
However, it was checked that the results did not change, at
the level of detail considered in this work, when this term was used.

Further simplifications can be made to the ionization matrix
elements using Eq. (\ref{EqFinalStation}a)
\begin{equation}\label{EqIonize}
	 \langle\phi^{(g)}_{\bf n};{\mathbf k}_0,t_b|
	\widehat V_L(t_b)|\Psi_i\rangle =
	 \langle\phi^{(g)}_{\bf n};-{\bf A}(t_b),t_b|
	\widehat V_L(t_b)|\Psi_i\rangle.
\end{equation}
For all $t_b$, the final continuum wave is 
$\langle {\bf r}|(-{\bf A}(t_b)),t_b\rangle = 
(2\pi)^{-(3/2)} e^{i [0]\cdot{\bf r}} = constant$ 
[see Eq. (\ref{EqCanonicalBasis})].
Neglecting also angular excitation of the nuclei (rotational time scales
are much longer than the current time scales of interest),
the ionization matrix element becomes
\begin{equation}
	 \langle\phi^{(g)}_{\bf n};{\mathbf k}_0,t_b|
	\widehat V_L(t_b)|\Psi_i\rangle \propto
	\langle \phi^{(g)}_{\bf n} | \phi^{(i)}_{\bf n}\rangle  \cos(t_b)
\end{equation}
where $|\phi^{(i)}_{\bf n}\rangle$ are the nuclear states of the D$_2$ molecule
and $|\phi^{(g)}_{\bf n}\rangle$ are the nuclear state of the D$_2^+$ ion.
The overlap $\langle \phi^{(g)}_{\bf n} | \phi^{(i)}_{\bf n}\rangle$ 
can be recognized as Franck-Condon factors modulating the ionization step.

With these last considerations taken into account, 
Eq. (\ref{EqAfterStation}) becomes
\begin{widetext}
\begin{eqnarray} \label{EqAfterStation2}
	\langle\phi^{(u)}_{E,J};{\bf k}_f,t|\Psi(t)\rangle & \propto & 
	- \sum_{\bf n} 
	\sum_{j=S,L} 
	e^{i\varphi({\bf k}_0,t^{(j)}_b,t^{(j)}_c) } 
	a_p(I_{p,{\bf n}},t^{(j)}_c-t^{(j)}_b)
	a_i(I_{p,{\bf n}},t^{(j)}_b)
	\\ \nonumber  & \times &
	\cos(t^{(j)}_b)
	\langle \phi^{(u)}_{E,J};{\bf k}_f|
	\widehat V_{ee}
	|\phi_{\bf n};-{\mathbf A}(t^{(j)}_b)\rangle  
	\langle \phi^{(g)}_{\bf n} | \phi^{(i)}_{\bf n}\rangle
	\\ \nonumber &\equiv& 
	\langle\phi^{(u)}_{E,J};{\bf k}_f| \widehat S_{DR} |\Psi_i\rangle,
\end{eqnarray}
\end{widetext}
where $\widehat S_{DR}$, defined by Eq. (\ref{EqAfterStation2}) 
is the approximate 
dissociative recollision scattering operator, and the (now extraneous) 
time-dependence of the electronic plane wave basis was dropped.
The D$^{+}$  kinetic energy spectrum $W(E)$ is found by integrating over 
the final scattering electron and angular momentum states:
\begin{equation}\label{EqSFWE}
	W(E) = 
    \sum_{J=1,3,5...}
	 \int d{\bf k}_f \left|
	 \langle\phi^{(u)}_{E,J};{\bf k}_f| \widehat S_{DR} |\Psi_i\rangle
     \right|^2 .
\end{equation}
The equivalent expression when starting with an initial 
superposition of D$_2$ states $\sum_i C_i |\Psi_i\rangle$ is
\begin{equation}\label{EqSFWE2}
	W(E) = 
    \sum_{J=1,3,5...}
	 \int d{\bf k}_f \left| \sum_i C_i
	 \langle\phi^{(u)}_{E,J};{\bf k}_f| \widehat S_{DR} |\Psi_i\rangle
     \right|^2 .
\end{equation}
Of interest is also the total yield defined as
\begin{equation}\label{EqWT}
	W_T = \int W(E)dE.
\end{equation}

Interfering pathways that can be used for coherent control
can be seen in Eqs. (\ref{EqAfterStation2}) and (\ref{EqSFWE2}).
The sum over ${\bf n}$ in Eq. (\ref{EqAfterStation2}), that is,
the vibrational motion of the nuclei between the moments of
ionization and recollision, offers
a means of pump-dump control, and also provides the opportunity
to probe the vibrational motion of the nuclear wave packet
excited to the $\Sigma_u$ state following ionization, 
the focus of the experiment \cite{CorkumRecollision}.
The pump-dump perspective is modeled and discussed in 
Sec. \ref{SecPumpDump}, the results of which are in excellent 
agreement with the
experiment, thereby confirming the validity of the SFA model.
The sum over $i$ in Eq. (\ref{EqSFWE2}) provides a means of
control through preparation of an initial vibrational
superposition state before the strong field is applied.
This control is similar in spirit to traditional bichromatic
control schemes, and is explored in Sec. \ref{SecBichromatic}.
There are also interferences arising from the so-called long and short
trajectories [sum over $j$ in Eq. (\ref{EqAfterStation2})].
However, these interferences lead to rapid oscillations of the
final yield as a function of ${\bf k}_f$, and average to zero
once the yields are integrated over ${\bf k}_f$.  

Note that in this formulation the momentum
${\bf k}$, referred to as the free electron momentum, is in
reality the relative momentum of the ion and the free electron,
and that the center-of-mass momentum was assumed to be zero.
In typical strong field scenarios, this subtlety is irrelevant.
However, it is important to stress this point 
in the present study since the field-free scattering formalism,
presented in the following section, will make use of both
the lab frame coordinates of the electron and ion as well
as the relative and center-of-mass coordinate.


\subsection{Field-Free Wave-Packet Scattering}
\label{SecFieldFree}

Consider then scattering in the absence of an external field.
We work within the $S$-matrix formalism \cite{Child,Taylor}
\begin{equation}
	S_{\bf ab} = 
	\left \langle {\bf b} \left |
	e^{-i\int_{-\infty}^{\infty}{\widehat H}dt}
	\right | {\bf a} \right \rangle
	= \delta_{\bf ab} - i T_{\bf ab}
\end{equation}
to calculate the transitions from the initial state
$|{\bf a}\rangle = |\phi^{(g)}_{\bf n};{\bf p}_i;{\bf P}_i\rangle$ to final 
state $|{\bf b}\rangle = |\phi^{(u)}_{E,J};{\bf p}_f;{\bf P}_f\rangle$,
where ${\bf p}_i$ (${\bf p}_f$) is the initial (final)
momentum of the scattering electron, and 
${\bf P}_i$ (${\bf P}_f$) is the initial (final)
momentum of the ion, both in the laboratory frame.
Within the Born approximation,
the transition matrix elements $T_{\bf ab}$ 
to first order in the electron-electron interaction are given by
\begin{eqnarray} \label{EqTab}
	T_{\bf ab} &=& \int dt'
	\left \langle{\bf a} \left |
	e^{-i\int_{t'}^{\infty}{\widehat H}_{\bf a}dt}
	\widehat V_{ee}
	e^{-i\int^{t'}_{-\infty}{\widehat H}_{\bf b}dt}
	\right | {\bf b} \right \rangle \\ \nonumber 
	&=& (2\pi) \delta(E_{\bf b}-E_{\bf a})
	\:\delta({\bf K}_f - {\bf K}_i) 
	 \left \langle \phi^{(u)}_{E,J};{\bf k}_f \left |
	\widehat V_{ee}
	\right | \phi^{(g)}_{\bf n};{\bf k}_i \right \rangle,
\end{eqnarray}
with energies 
\begin{subequations}
\begin{eqnarray}
	E_{\bf a} &=& \frac{{\bf p}_i^2}{2} + \frac{{\bf P}_i^2}{2m_I}
	+ E_{\bf n}, \\
	E_{\bf b} &=& \frac{{\bf p}_f^2}{2} + \frac{{\bf P}_f^2}{2m_I}
	+ E.
\end{eqnarray}
\end{subequations}
Here ${\bf K}_j$ and ${\bf k}_j$ ($j=i,f$) are the center-of-mass
and relative momenta of the e-D$_2^+$ system
\begin{subequations}\label{EqLabRelative}
\begin{eqnarray}
	{\bf K}_j &=& {\bf p}_j  + {\bf P}_j, \\
	{\bf k}_j &=& \frac{m_I {\bf p}_j - {\bf P}_j}{m_I+1}, 
\end{eqnarray}
\end{subequations}
and $m_I = 2 m_p$ is the mass of the ion.  

Due to the $\delta({\bf K}_f - {\bf K}_i)$ term in Eq. (\ref{EqTab}),
which represents conservation of total momentum,
${\bf K}_f = {\bf K}_i \equiv {\bf K}$ is a constant of motion.  
The scattering problem is then 
solved for each ${\bf K}$ individually, after which the yields
are averaged incoherently over the ${\bf K}$ distribution.
First, the projection of the scattered wave function onto
the final state basis is calculated as
\begin{widetext}
\begin{eqnarray}\label{EqTemp}
	|\langle \phi^{(u)}_{E,J};{\bf k}_f;{\bf K}|\Psi_f\rangle |^2
	&=& (2\pi)^2 \Bigg |\sum_{\bf n}\int d{\bf k}_i
	\delta(E_{\bf b} - E_{{\bf a}}) 
	\\ \nonumber & \times &
	\left \langle \phi^{(u)}_{E,J};{\bf k}_f \left |
    \widehat V_{ee}
    \right | \phi^{(g)}_{\bf n};{\bf k}_i \right \rangle
	\langle \phi^{(g)}_{\bf n};{\bf k}_i;{\bf K} |\Psi_0\rangle\Bigg|^2.
	\\ \nonumber
\end{eqnarray}
With the initial translational 
momenta of both the electron and ion antiparallel in the 
laboratory frame, as done in the following sections, 
Eq. (\ref{EqTemp}) can be reduced to 
\begin{eqnarray}
	|\langle \phi^{(u)}_{E,J};{\bf k}_f;K|\Psi_f\rangle |^2
	&=& (2\pi)^2 
    \left|
     \sum_{\bf n}
    (1/k_{i 0}) 
	\left \langle \phi^{(u)}_{E,J};{\bf k}_f \left |
    \widehat V_{ee}
    \right | \phi^{(g)}_{\bf n};k_{i0} \right \rangle
    \left\langle \phi^{(g)}_{\bf n};k_{i0} ; K|
    \Psi_0 \right \rangle  
	\right|^2,
\end{eqnarray}
\end{widetext}
where
\begin{equation}
	k_{i 0} = \sqrt{2\left(|{\bf k}_f|^2/2+E
	- E_{\bf n} \right)}.
\end{equation}
The partial yields of interest are calculated
by integrating over the relevant final states
\begin{equation}
	W_K(E) = 
    \sum_{J=1,3,5...}
	 \int d{\bf k}_f \left| 
	\langle \phi^{(u)}_{E,J};{\bf k}_f; K|\Psi_f\rangle 
     \right|^2
\end{equation}
\begin{equation}
	W(E) = \int W_K(E) dK, 
\end{equation}
and the total yield is  given by Eq.(\ref{EqWT}).

For the case of $e + D_2^+$ collisions, the electron impact matrix elements
$\left \langle \phi^{(u)}_{E,J};{\bf k}_f \left |\widehat V_{ee}
    \right | \phi^{(g)}_{\bf n};k_{i0} \right \rangle$
are computed as previously described (Ref. \cite{Me}).

\section{Control Scenarios}

\subsection{Vibrational Probing and Pump-Dump Control}
\label{SecPumpDump}

Consider first the original experiments of Niikura {\it et al.} 
\cite{CorkumRecollision}, where the D$^+$  kinetic energy 
spectrum was measured as a function of laser wavelength.
The basic idea was to exploit the frequency-dependence of
the time delay between the moment of ionization $t_b$ and moment 
of collision $t_c$, implicit in Eq. (\ref{EqFinalStation}b),
in order to probe the vibrational motion of the D$_2^+$ vibrational 
wave packet.  This vibrational probing scenario can
readily be identified as a pump-dump control scenario:
the ionization step pumps the nuclei to the $\Sigma_g$ surface
while the recollision dumps the nuclear wave packet 
to the dissociative surface.  Varying the laser frequency
controls the pump-dump time delay.  In fact, the vibrational
probing experiments were guided by this close analogy with
pump-dump control.

\begin{figure}[t!]
	\centering
	\includegraphics[width=\columnwidth]{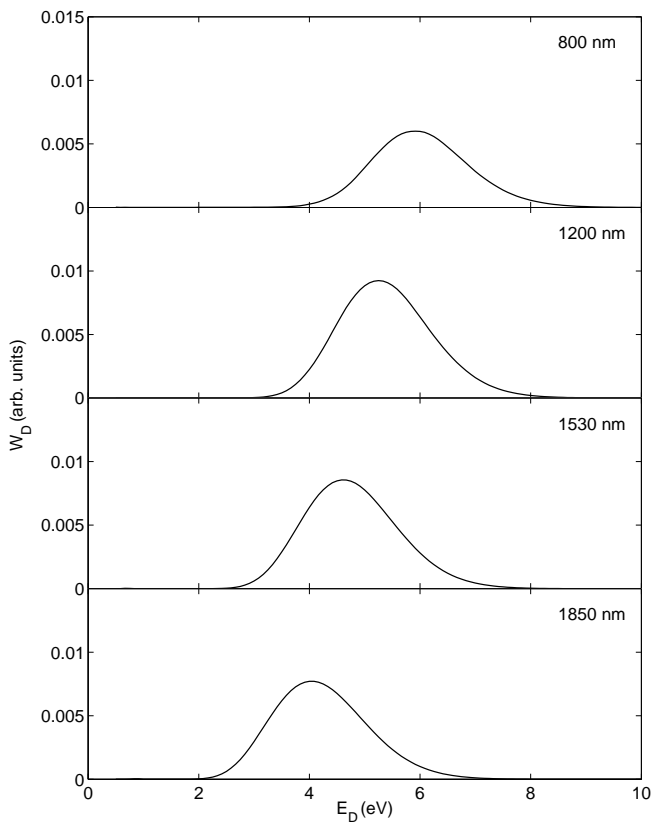}
	\caption{Pump-dump control of the dissociative recollision
	D$^+$  spectrum.  The ionization event
	provides the 'pump', while recollision provides the 'dump'.
	The pump-dump time delay is controlled by varying the driving
	frequency.}
	\label{FigPumpDump}
\end{figure}

Figure \ref{FigPumpDump} demonstrates the vibrational probing/pump-dump 
scenario using the SFA model.  Here, and through the rest of the paper,
we plot the D$^+$  kinetic energy $E_D$ spectrum, 
$W_D(E_D) \equiv 2W(E/2)$, using Eq. (\ref{EqSFWE}).
The factors of 2 arise from the fact that $E$ is the relative
energy shared by the D and D$^{+}$; since the D$^+$ are much
heavier than the scattered electron, the center-of-mass energy
of the two D$^+$ is negligible and the relative energy $E$ is 
shared equally between the kinetic energies of the D$^+$.
Here $D_2$ was initially in its ground vibrational and 
rotational state, and 
the field strength used was ${\cal E}_0 = 0.065$ au 
($I_0 = 1.4\times10^{14}$ W/cm$^2$).  This field strength
is used throughout the paper.
The results were calculated for a single recollision event.
Multiple recollisions would increase the absolute yield, but would
not change the relative yields across the spectrum.
The four panels show the D$^+$ energy spectrum 
for the four different wavelengths (800, 1200, 1530, and 1850 nm) 
used in Ref. \cite{CorkumRecollision}.  The simulated
results are in excellent agreement with the experiment 
\cite{CorkumRecollision}, thus
validating the SFA model.  $W_D(E_D)$ clearly
depends on the wavelength, and this dependence reflects the motion
of the D$_2^+$ nuclear wave function on the $\Sigma_g$ surface
between the moments of ionization $t_b$ and recollision $t_c$, 
both providing a means to probe the vibrational motion as well as
offering an example of pump-dump control.  

\subsection{Bichromatic Coherent Control}
\label{SecBichromatic}

The orthodox bichromatic coherent control scenario \cite{BrumerShapiro} 
utilizes CW lasers and involves first creating 
an initial superposition state and then driving the superposition
to a final state where the components of the superposition can interfere.
This scenario can be extended to the time domain, and 
to attosecond dissociative recollision, in particular 
by creating an initial vibrational superposition in D$_2$.
Since the recolliding electron has a broad bandwidth (energies
of recollision vary from 0 to 3.17 $U_p$ \cite{Plasma}, where 
$U_p = ({\cal E}_0/\omega)^2/4$ is
the pondermotive potential), the various initially-populated
vibrational states will overlap in the nuclear continuum on the
antibonding surface following collisional excitation, and make
possible the bichromatic control of the D$^+$  energy spectrum.

\begin{figure}[t!]
	\centering
	\includegraphics[width=\columnwidth]{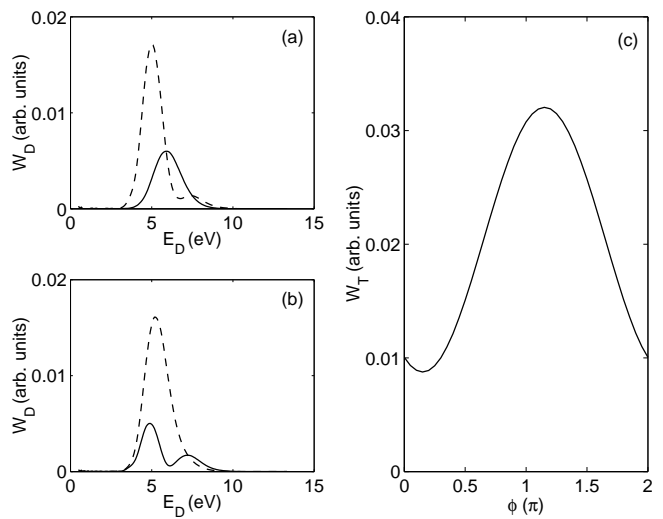}
	\caption{Single-cycle bichromatic control of dissociative recollision.
	D$^+$ spectrum for initial (a) $|0\rangle$ vibrational
	state (solid), and initial $|1\rangle$ state (dashed);
	(b) initial $|0\rangle+|1\rangle$ superposition (solid)
	and $|0\rangle-|1\rangle$ superposition (dashed).
	(c) D$^+$ yield from $|0\rangle+\exp(i\phi)|1\rangle$
	state as a function of $\phi$.}
	\label{FigBichromaticCalc}
\end{figure}

This scenario is illustrated in Fig. \ref{FigBichromaticCalc}
using a superposition of the $|\nu\rangle = |0\rangle$ 
and $|1\rangle$ vibrational states of D$_2$, with rotation in 
the ground state.  The results correspond to a single 
recollision event and a driving field of frequency 
$\omega = 0.0569$ au (800 nm).  Panel (a) plots the
final D$^{+}$  spectra starting from $|0\rangle$ (solid) and $|1\rangle$
(dashed) individually, while panel (b) plots the spectra arising 
from the vibrational superpositions $|+\rangle = |0\rangle+|1\rangle$
(solid) and $|-\rangle = |0\rangle-|1\rangle$ (dashed).
Panel (c) plots the total integrated D$^{+}$  yield $W_T$ for the
vibrational superposition state 
$|\phi\rangle = |0\rangle+e^{i\phi}|1\rangle$ as $\phi$ is varied.
The range of control is significant.
These results should strongly motivate an experimental 
study of bichromatic coherent control
via strong field dissociative recollision: the D$^{+}$ 
spectrum and total yield can be actively controlled by using 
phase-coherent initial states.

\begin{figure}[t!]
	\centering
	\includegraphics[width=\columnwidth]{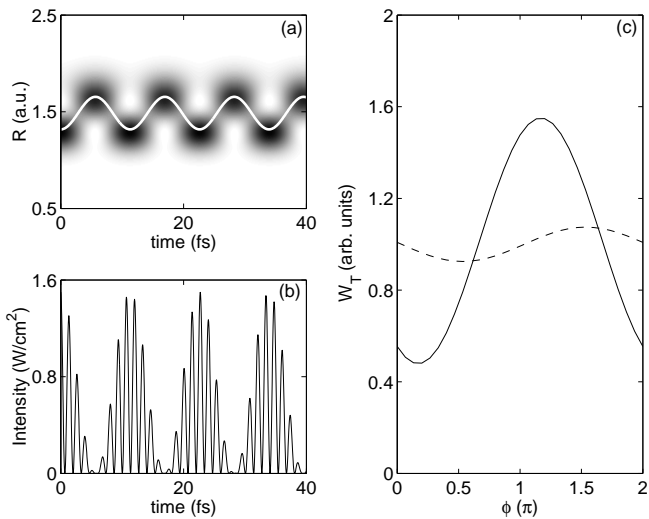}
	\caption{Many-cycle bichromatic control of dissociative recollision.
	(a) Density of D$_2$ vibrational wave function as a function
	of time.  The undulation of the density is highlighted by the
	superimposed wave. (b) Incident laser intensity as a function of time
	for the two-color field of Eq. (\ref{EqE2}) with ${\cal E}_0 =
	$, $\omega = 0.0569$ au, and $\Delta \omega = 0.0135$ au
	(c) Control of D$^+$ yield as a function of $\phi$ for a
	monochromatic incident field of frequency $\omega$ (dashed) and
	the two-color field (solid).}
	\label{FigBichromaticMany}
\end{figure}

As in the previous section, these results include only
a single recollision event, thereby effectively simulating the result
of driving the recollision with a single-cycle pulse.  In the
case of pump-dump control, adding more cycles did not change
the resulting D$^{+}$  spectrum.  However, this is not the case
for the bichromatic scenario.  If the number of cycles is
large enough so that the total time duration of the driving
pulse is larger than the vibrational period of the
initial superposition state, then control will disappear,
since each cycle would contribute to the final D$^{+}$ 
spectrum with a different phase $\phi$ due to the evolution
of the vibrational eigenstates from one cycle to another.
However, control can be restored in the many cycle regime by 
using pulses at two different frequencies, 
\begin{eqnarray}\label{EqE2}
	{\bf E}_2(t) &=& ({\cal E}_0/2)
	[\cos(\omega_1 t) +\cos(\omega_2 t)]\hat x
	\nonumber \\
	 &=& {\cal E}_0\cos((\Delta\omega/2) t) \cos(\omega t)\hat x,
\end{eqnarray}
where $\omega_1 = \omega + \Delta\omega/2$,
$\omega_2 = \omega - \Delta\omega/2$,
and $\Delta\omega$ is set equal to the energy difference between the 
two states of the initial superposition.
In this case, the two frequencies will cause beats in the
envelope of the long pulse that are timed to the motion of
the internal state superposition (see Figs. 
\ref{FigBichromaticMany}a and b).  
Here the central frequency is set to $\omega = 0.0569$ au (800 nm).
Varying the relative phase of the superposition, or alternatively 
varying the relative phase of the two frequencies, then 
varies the relative timing between the internal state motion
and the envelope beats, and allows one to control which phases
of the vibrational superposition contribute to the
recollision events.

Figure \ref{FigBichromaticMany}c shows the control, as a function $\phi$,
when using a monochromatic field (dashed) and the two-color field 
[Eq. (\ref{EqE2})] with $\omega = 0.0596$ au and 
$\Delta \omega = E_{\nu=1} - E_{\nu=0} = 0.0135$ au
The calculations included eleven cycles of the 800 nm
carrier, corresponding to two vibrational periods of the
D$_2$ superposition.
When solving the stationary phase equations for the two-color 
laser field, the field and vector potential were treated as 
pure sine waves over the half cycle,
\begin{equation}
	{\bf E}_2(t) \approx
	 {\cal E}_0\cos((\Delta\omega/2) T_n) \cos(\omega t)\hat x
\end{equation}
and
\begin{equation}
	{\bf A}_2(t) \approx -\frac{{\cal E}_0}{\omega} 
	\cos((\Delta\omega/2) T_n)\sin(\omega t) \hat x,
\end{equation}
where the amplitude of the beat envelope $\cos((\Delta\omega/2) T_n)$
was evaluated at the time $T_n$ corresponding to the peak 
field strength of the half cycle, defined 
by $\cos(\omega T_n) = 1$.   For ease of comparison, both 
control plots were normalized
so that the average yield as a function of $\phi$ is unity
to compensate for the different ionization yield that depends 
exponentially on the instantaneous electric field strength
[Eq. (\ref{EqAiry})].
As can be seen from Fig. \ref{FigBichromaticMany}c,
the control with the monochromatic wave is greatly diminished
for many cycles, and the control eventually reaches zero in
the limit of infinite number of cycles.  However, control
in the case of the two-color field persists, and does not
diminish further when additional cycles are added.

\subsection{Coherently Controlled Reactive Scattering}

An interesting question arises when one considers dissociative 
recollision from the perspective of coherently controlled
reactive scattering.  Traditional (non-control) 
scattering scenarios \cite{Child,Taylor}
are almost all conducted in a field-free environment and use nearly
monoenergetic beams.  However, the present scenario necessarily
involves the presence of a strong laser field and wave packets
of translational momentum.  Coherent control of reactive scattering 
in the continuous
beam regime \cite{BrumerScattering} is now well understood, and the 
wave packet extension has recently been explored 
\cite{Me}.  Although dissociative recollision clearly
involves a scattering process, it is also coupled to the ionization
step and the in-field propagation, and hence it is not 
entirely clear to what extent
it can be considered as a true example of coherently controlled
reactive scattering.  The easiest way to address this issue
is to compare the strong field recollision scenario to its
$D_2^++e^-$ field-free counterpart using the field-free scattering
formalism presented in Sec.\ref{SecFieldFree}.

The incident scattering state used to mimic the recollision
event has the form
\begin{equation}\label{EqPsiTrans}
	|\Psi_0\rangle = \sum_{\bf n} \int d{\bf K}d{\bf k} \: 
	                              a_{\bf n}\psi({\bf k},{\bf K}) 
	                              |\phi^{(g)}_{\bf n},{\bf k},{\bf K}\rangle,
\end{equation}
where $a_{\bf n}$ and $\psi({\bf k},{\bf K})$ are the vibrational 
and translational expansion coefficients respectively.
In the recollision scenario, the collision between the electron
and the ion lasts about 0.5 fs. For
the laser intensity used above, at 800 nm, the peak recollision
momentum is $p_{max} = \sqrt{2\times 3.17 U_p} \approx 1.5$ au.
This recollision wave packet can then be adequately represented 
in the field-free case by the following translational superposition
\begin{eqnarray}\label{EqMomentumCoeffs}
	\psi({\bf k},{\bf K}) &=& \frac{1}{\sqrt{\pi \Delta K \Delta k}}
	\exp\left[ -\frac{1}{2} \left( 
	            \frac{{\bf K}-{\bf K}_0}{\Delta K} \right)^2 \right]
	\nonumber \\ & \times &
	\exp\left[ -\frac{1}{2} \left( 
	            \frac{{\bf k}-{\bf k}_0}{\Delta k} \right)^2 \right],
\end{eqnarray}
with relative coordinate parameters given by 
$\Delta k = 0.5$ au and ${\bf k}_0$ = 1.5 au, parameters chosen
to mimic the recollision wave packet observed in the simulations
of the previous sections.   Since the scattering is independent 
of the center-of-mass parameters, any values 
of $\Delta K$ and ${\bf K}_0$ can be used.  Here we set 
these parameters to $\Delta K = 1.0$ au and ${\bf K}_0$ = 0 au,
where ${\bf k}_0$ and ${\bf K}_0$ are parallel to each other.
In order to match the D$_2^+$ vibrational wave packet created in
the strong field case, the vibrational coefficients are set to
\begin{equation}\label{EqVibCoeffs}
	a_{\bf n} \propto \langle \phi^{(g)}_{\bf n}|\phi^{(i)}_0\rangle 
	\exp\left[-\frac{1}{3}
	          \frac{(2I_{p,{\bf n}})^{3/2}}
	{\left|{\mathbf E}(t_b) \right|} - iE_{\bf n} \tau_d \right]
\end{equation}
where $\langle \phi^{(g)}_{\bf n}|\phi^{(i)}_0\rangle$ are the
Franck-Condon overlaps, and the exponential mimics both the 
ionization amplitude [see Eq. (\ref{EqAiry})] and the time delay 
$\tau_d$ = 2.0 fs
between the momentum of ionization and recollision ($\tau_d$
is $\sim$3/4 of the laser cycle \cite{Plasma}).  
This distribution of vibrational states on the $D_2^+$ surface 
follows from the above presented SFA theory by evaluating
Eq. (\ref{EqAfterStation2}) after ionization but before recollision,
and has been observed experimentally \cite{Urbain}.
Crucial from the perspective of coherent control is that
the combination of a spread in population of vibrational states
$|\phi^{(g)}_{\bf n}\rangle$ [Eq. (\ref{EqVibCoeffs})] and spread 
in incident momentum [Eq. (\ref{EqMomentumCoeffs})] leads to multiple 
energetically degenerate pathways to product states.

\begin{figure}[t!]
	\centering
	\includegraphics[width=\columnwidth]{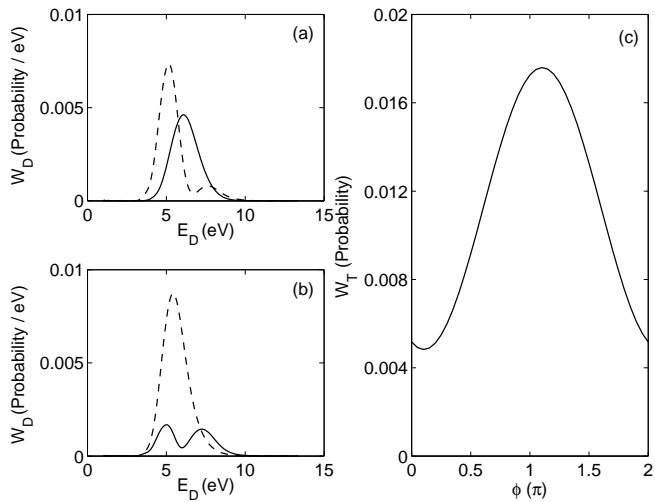}
	\caption{Field-free wave-packet coherent control of
	electron impact dissociation.  
	D$^+$ spectrum for initial (a) $|0\rangle$ vibrational
	state (solid), and initial $|1\rangle$ state (dashed);
	(b) initial $|0\rangle+|1\rangle$ superposition (solid)
	and $|0\rangle-|1\rangle$ superposition (dashed).
	(c) D$^+$ yield from $|0\rangle+\exp(i\phi)|1\rangle$
	state as a function of $\phi$.}
	\label{FigScattering}
\end{figure}

Figure \ref{FigScattering} plots the resulting D$^{+}$  yields for 
the field-free scattering.  Upon comparing panels (a) and (b) of
Figs. \ref{FigScattering} and \ref{FigBichromaticCalc}, one sees
that the D$^{+}$  spectra resulting from the strong field and field-free
scenarios are essentially the same.  Small variations are present,
for example the distributions in  the field-free case are all
a bit narrower than their field-free counterparts, but this
is likely due to small differences between the actual recollision
state and that used to mimic this state in the field-free scenario.  
When comparing the phase dependence of the D$^{+}$ 
yields [panel (c) of Figs. \ref{FigScattering} and 
\ref{FigBichromaticCalc}], one sees that the general sinusoidal 
dependence is perfectly captured by the field-free scenario.
It is therefore clear that the strong field 
and field-free cases behave in essentially similar ways, 
and hence that the dissociative recollision 
scenario offers an example of wave-packet control of reactive 
scattering.

\subsection{Entanglement in Dissociative Recollision}

We now address the importance of entanglement in dissociative
recollision.  Specifically, the literature suggests a connection between
the sub-femtosecond
time resolution present in the vibrational probing of the D$^+_2$
wave packet and entanglement \cite{CorkumRecollision,Entangle1,EntangleMarkus}.
The entanglement under consideration is that between the electron and ion
after ionization and before recollision:  
when the D$_2$ is ionized, the resultant wave function is separable in 
relative and center-of-mass coordinates, similar to Eq. (\ref{EqPsiTrans}),
but not separable in the ion and electron coordinates.
The question of entanglement is also central to the coherent 
control of reactive scattering \cite{BrumerScattering,Me}.  In that
case, initial state 
entanglement is required for control when using superpositions
of a few isolated translational plane waves 
\cite{BrumerScattering}.  Recent results show, however, that
initial state 
entanglement is not necessary when working with specific scenarios
utilizing wave packets of translational
motion \cite{Me} involving  energetically degenerate pathways.  
The dissociative recollision scenario features
such pathways, as noted above, and hence it is conceivable that
entanglement is neither crucial for control nor for vibrational probing.
This is examined below.

\begin{figure}[t!]
	\centering
	\includegraphics[width=\columnwidth]{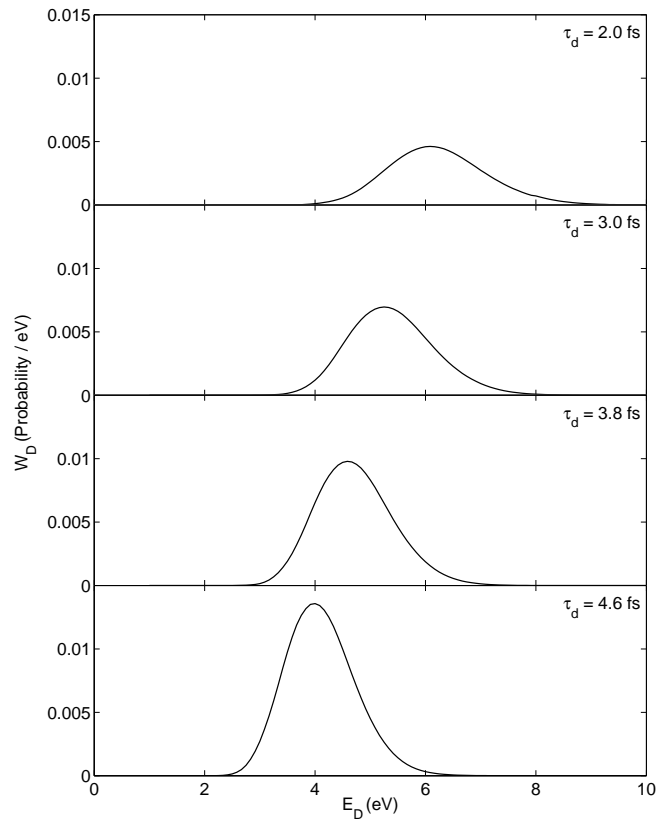}
	\caption{Non-entangled wave packet probing of vibrational 
	motion.  To be compared with Fig. \ref{FigPumpDump}.
	The $\tau_d$ values correspond to the pump-dump time delays
	for the wave lengths used in Fig.\ref{FigPumpDump}.}
	\label{FigNoEntangleProbe}
\end{figure}

To gain insight into the role of entanglement, we consider once again
the analogous field-free case with independently prepared
wave packets of D$_2^+$ and e$^-$, but this time without entanglement
present.  That is, our initial state is
\begin{equation}\label{EqPsiTrans2}
	|\Psi_0\rangle = \sum_{\bf n} \int d{\bf P}d{\bf p} \: 
	                              a_{\bf n}\psi({\bf p},{\bf P}) 
	                              |\phi^{(g)}_{\bf n},{\bf p},{\bf P}\rangle,
\end{equation}
where
\begin{eqnarray}
	\psi({\bf p},{\bf P}) &=& \frac{1}{\sqrt{\pi \Delta P \Delta p}}
	\exp\left[ -\frac{1}{2} \left( 
	            \frac{{\bf P}-{\bf P}_0}{\Delta P} \right)^2 \right]
	\nonumber \\ & \times &
	\exp\left[ -\frac{1}{2} \left( 
	            \frac{{\bf p}-{\bf p}_0}{\Delta p} \right)^2 \right],
\end{eqnarray}
the electronic parameters are given by 
$\Delta p = 0.5$ au and ${\bf p}_0$ = 1.5 au, and the ionic
parameters are  $\Delta P = 1.0$ au and ${\bf P}_0$ = 0.
The $a_{\bf n}$ are the same as those used above.
Recall that the ${\bf P}_0$ and ${\bf p}_0$ are momenta in
the laboratory frame.

Figure \ref{FigNoEntangleProbe} shows the results of mimicking the
dissociative recollision probing/pump-dump scenario using
the field-free non-entangled scattering scenario.  
The 'pump-dump time delay' was varied by controlling $\tau_d$, 
and results for the four values of $\tau_d$ corresponding to the 
wave lengths used in the strong field pump-dump scenario are
shown.  Upon comparing Fig. \ref{FigNoEntangleProbe} with 
Fig. \ref{FigPumpDump}, one sees that the non-entangled
field-free scattering scenario does capture the vibrational
motion just as well as did the strong field recollision scenario.

Note that the $W_D(E_D)$ peak height increases as $\tau_d$ 
becomes larger in the field-free scenario, but the
analogous peak height in the strong field case remains roughly constant. 
The change in total yield in the field-free case is present
because the cross section for electron impact dissociation of D$_2^+$
depends strongly on the internuclear distance; as the bond is
allowed to stretch, it becomes easier to dissociate.
In the strong field case, as the wave length is varied and
the pump-dump time delay increases,
the continuum electron wave packet spreads more 
in the transverse direction, thus counteracting the expected increase
in the total yield as the bond length stretches.

These results directly imply that it is the temporal correlations 
between the incident electron wave packet and the vibrational
motion of the ion that is necessary for probing/control, 
but entanglement is not.  Further, the 
general agreement between Fig. \ref{FigNoEntangleProbe} and
Fig. \ref{FigPumpDump} strengthens the argument that 
dissociative recollision scenarios can be considered
as examples of wave-packet coherent control of reactive scattering.
The theory of wave-packet coherent control of reactive
scattering is discussed in detail in Ref.\cite{Me}.

\begin{figure}[t!]
	\centering
	\includegraphics[width=\columnwidth]{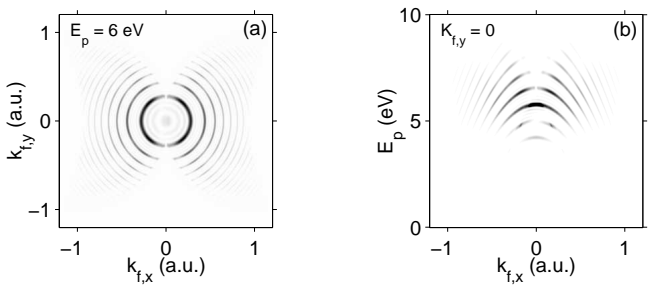}
	\caption{Signatures of entanglement in the multiparticle
	distributions.  (a) Electron momentum spectrum, for
	$E_D$ = 6 eV, showing ATI structure. (b) Coincident 
	D$^+$ /electron spectrum, for $k_{f,y}$ = 0, showing 
	ATI-like structures in the D$^+$ kinetic energy distribution.
	Both spectra are taken at $k_{f,z}$=0.}
	\label{FigDetectEntangle}
\end{figure}

Although electron-ion entanglement is not central to probing or control, the 
fact remains that entanglement is present.  The entangled 
nature of the multiparticle density can be revealed through
coincidence measurements, where the D$^+$  kinetic energy spectrum
is measured together with the recollision electron.
The ionization, and subsequent recollision, certainly leads
to multiparticle correlations in the outgoing state.  However, 
to demonstrate entanglement, one must show non-separable wave 
function characteristics, which include correlated observables 
and coherence.  When using a driving pulse with many cycles, the
outgoing flux originating from each half cycle will interfere,
hence providing a means to verify the coherence of the
final wave function.  

Figure \ref{FigDetectEntangle}a shows
a slice through the final momentum distribution of the 
scattered electron taken at $E_D$ = 6 eV and $k_{f,z}$ = 0, 
calculated for a five-cycle 800 nm pulse.  The ring structure, similar to
the well known Above-Threshold Ionization (ATI) 
peaks, appear as result of the periodic 
nature of strong field ionization.  Each half-cycle 
an electron wave packet is launched into the continuum.
All these wave packets then interfere to produce the rings.
In an energy representation each ring is 
separated by one unit of the photon energy.  
Since dissociative recollision also occurs 
each half cycle, similar ATI-like structures exist
in the D$^{+}$  energy spectrum, Fig.\ref{FigDetectEntangle}b,
again a result of interferences between the outgoing
D$^{+}$  flux originating from each half-cycle.
The spectrum shown in Fig. \ref{FigDetectEntangle}b corresponds
to $k_{f,y}$ = 0 and $k_{f,z}$ = 0.
Unlike one-electron ATI, the ring structures resulting
from dissociative recollision only appear when both
the recollision electron and the D$^{+}$  are measured
in coincidence, that is, the structure exist only in such
spectra.  Further, since the rings are a result of quantum
interference, the resulting  wave function is
necessarily coherent.  Observation of these ATI-like rings
in the coincidence spectra would then demonstrate 
the entanglement that is necessarily present during the 
dissociative recollision process.

\section{Summary}

We have considered the attosecond dissociative 
electron recollision in D$_2$ from the perspective of
coherent control.  Both pump-dump and bichromatic control
schemes were demonstrated in this system.  Further,
by direct comparison of the strong field recollision
scenario with analogous field-free cases, it was shown
that this system offers an example of wave-packet coherent control
of reactive scattering.  By constructing and analyzing similar 
scenarios involving scattering of non-entangled field-free wave 
packets, it was shown that the electron-ion entanglement effectively plays
no role in the control scenario presented herein
nor in the closely related problem of recollision-based
sub-femtosecond vibrational probing \cite{CorkumRecollision}.
Finally, detectable signatures of electron-ion entanglement in the outgoing
multiparticle states, arising from temporal interferences,
were identified.

\section{Acknowledgments}
M.S. would like to thank O. Smirnova and M. Ivanov
for numerous insightful discussions.
Both authors acknowledge positive discussions with P. Corkum.  
This work was supported by the National Science
and Engineering Research Council of Canada.

\appendix



\begin{thebibliography}{}
 
\bibitem{PumpDump} 
D.J. Tannor, S.A. Rice, J. Chem. Phys. {\bf 85}, 5013 (1985); D.J.
Tannor, R. Kosloff, S.A. Rice, J. Chem. Phys. {\bf 85}, 5805 (1985).

\bibitem{BrumerShapiro}
M. Shapiro and P. Brumer, \textit{Principles of the Quantum
Control of Molecular Processes},
(John Wiley \& Sons, New Jersey, USA, 2003).

\bibitem{STIRAP}
K. Bergmann, H. Theuer, and B. W. Shore, Rev. Mod. Phys. 
{\bf 70}, 1003 (1998).

\bibitem{BrumerScattering}
M. Shapiro and P. Brumer, Phys. Rev. Lett. {\bf 77}, 2574 (1996); 
A. Abrashkevich, M. Shapiro, and P. Brumer, 
Phys. Rev. Lett. {\bf 81}, 3789 (1998);
82, 3002(E) (1999); 
Chem. Phys. 267, 81 (2001);
P. Brumer, A. Abrashkevich, and M. Shapiro, 
Discuss. Faraday Soc. 113, 291 (1999).

\bibitem{Me}
M. Spanner and P. Brumer, Phys. Rev. A (preceding paper).

\bibitem{Plasma} 
P. B. Corkum, Phys. Rev. Lett. {\bf 71}, 1994 (1993).

\bibitem{HHG}
M. Lewenstein, Ph. Balcou, M.Yu. Ivanov, Anne L'Huillier, and P.B. Corkum,
Phys. Rev. A {\bf 49}, 2117 (1994).

\bibitem{Diffraction}
M. Lein, J.P. Marangos, and P.L. Knight,
Phys. Rev. A {\bf 66}, 051404 (2002);
M. Spanner, O. Smirnova, P.B. Corkum, and M.Yu, Ivanov
J. Phys. B: At. Mol. Opt. Phys. {\bf 37}, L243 (2004);
S.N. Yurchenko, S. Patchkovskii, I.V. Litvinyuk, P.B. Corkum,
and G.L. Yudin, Phys. Rev. Lett. {\bf 93}, 223003 (2004).

\bibitem{CorkumRecollision}
H. Niikura, F. L\'egar\'e, R. Hasbani, A.D. Bandrauk,
M.Yu. Ivanov, D.M. Villeneuve and P.B. Corkum,
Nature (London) {\bf 417}, 917 (2002);
H. Niikura, F. L\'egar\'e, R. Hasbani, M.Yu. Ivanov,
D.M. Villeneuve and P.B. Corkum,
Nature (London) {\bf 421}, 826 (2003).


\bibitem{foot1}
Indeed, NRC scientists such as Paul Corkum and Albert Stolow have
viewed all pump-probe experiments as examples of coherent control.

\bibitem{Entangle1}
J. Hu, K.-L. Han, and G.-Z. He, Phys. Rev. Lett. {\bf 95}, 123001 (2005).

\bibitem{EntangleMarkus}
M. Kitzler and M. Lezius, Phys. Rev. Lett. {\bf 95}, 253001 (2005).

\bibitem{Reiss}
H.R. Reiss, Prog. Quant. Electr. {\bf 16}, 1 (1992).

\bibitem{volkov} 
W. Gordon, Z. Phys. {\bf 40}, 117 (1926);
D.V. Volkov, Z. Phys. {\bf 94}, 250 (1935).

\bibitem{Coulomb}
O. Smirnova, M. Spanner, and M.Yu. Ivanov, 
 J. Phys. B: At. Mol. Opt. Phys. 39, S307 (2006).

\bibitem{Keldysh}
L.V. Keldysh, Zh. eksp. teor. Fiz. {\bf 47}, 1945 (1964)
[English translation: Soviet Phys. JETP {\bf 20}, 1307 (1965)].

\bibitem{Child}
M.S. Child, {\it Molecular Collision Theory} (Academic Press, London,
UK, 1974). 

\bibitem{Taylor}
J.R. Taylor, 
{\it Scattering Theory: The Quantum Theory of Nonrelativistic Collisions} 
(John Wiley \& Sons, New York, NY, 1972).

\bibitem{Urbain}
X. Urbain, B. Fabre, E.M. Staicu-Casagrande, N. de Ruette, 
V.M. Andrianarijaona, J. Jureta, J.H. Posthumus, A. Saenz, E. Baldit,
and C. Cornaggia, Phys. Rev. Lett. {\bf 92}, 163004 (2004).

\bibitem{BunkinTugov}
F.V. Bunkin and I.I. Tugov, Phys. Rev. A {\bf 8}, 601 (1973).



\end{thebibliography}
\end{document}